# On the relation between microstructure and impact toughness of 17-4PH stainless steel produced by powder bed fusion laser beam (PBF-LB)

Renata De Oliveira MELO[a], Jean-Bernard VOGT[a], Eric NIVET[b], Flore VILLARET[c], Christophe GROSJEAN[b], Eric BAUSTERT[d], Nhu-Cuong TRAN[c], Jérémie BOUQUEREL[a,*], Gang JI[a,*]

[a]CNRS, INRAE, Centrale Lille, UMR 8207-UMET-Unité Matériaux et Transformations, Université de Lille, 59000 Lille, France

[b]CETIM, 52 Avenue Félix Louat, 60300 Senlis, France

[c]EDF R&D, Materials and Mechanics of Components Department, Av. les Renardieres, 77818 Moret sur Loing, France

[d]R&D MMB, Volum-e, 1 Chemin de la Fonderie, 76340 Blangy-sur-Bresle, France

*Corresponding authors: jeremie.bouquerel@centralelille.fr (Dr. Jérémie BOUQUEREL); gang.ji@univ-lille.fr (Dr. Gang JI)

## ABSTRACT

This work investigated the effects of aging heat treatments on the microstructure and, consequently, the quasi-static (tensile properties) and dynamic (impact toughness) mechanical behaviour of a 17-4PH stainless steel produced by powder bed fusion laser beam (PBF-LB). Multiscale microstructural characterization, using X-ray diffraction, scanning and transmission electron microscopy and electron backscatter diffraction, was conducted to establish quantitative correlations between microstructural evolution and mechanical performance, providing insight into the mechanisms governing plasticity and fracture. The PBF-LB specimens exhibited tensile strengths comparable to or exceeding those of conventionally manufactured counterparts but consistently showed significantly lower impact toughness, regardless of heat treatment conditions. Within the complex microstructure, strain-induced transformation of reversed austenite was found to enhance ductility and impact toughness. $SiO_2$ inclusions, originating from the starting powder, were identified as nucleation sites for micro-cavities and proved detrimental to impact toughness. Meanwhile, the distribution of Cu-rich nanoprecipitates could be tailored to favour either tensile strength through Orowan strengthening or impact toughness by enhancing local plasticity, but not both simultaneously. This work highlights the pronounced strength-toughness trade-off inherent in PBF-LB-produced 17-4PH alloys and reveals the interplay between strength, ductility, and toughness. These findings underscore the need for further research into dynamic loading mechanical properties, especially for demanding applications such as those in the nuclear sector.

# 1. Introduction

The development of advanced manufacturing processes is driving industrial modernization, enabling achievements once considered unattainable. In particular, the layer-by-layer approach of additive manufacturing (AM) allows for the creation of complex geometries that were previously impossible to produce using conventional metallurgical methods. This capability has become a practical reality across various industries. Additionally, AM offers the potential to transform maintenance strategies in industrial facilities by significantly reducing lead times [1]. Among AM techniques, powder bed fusion laser beam (PBF-LB) stands out for its design flexibility, ability to produce near-net-shape components, and suitability for fabricating intricate geometries [2]. Despite the maturity of AM as a technology, key questions remain regarding the process–structure–property relationships, particularly the mechanisms governing plastic deformation in metals fabricated by this method [3].

Industries are increasingly focusing on AM for the production of safety-critical components such as pressure vessels, complex piping systems, and load-bearing structures. These applications require an optimal balance between high strength and sufficient impact toughness. However, achieving this balance remains a challenge, as strength and toughness typically exhibit opposing trends. Materials with high strength often display reduced ductility and greater brittleness [4, 5]. This underscores the need for detailed microstructural studies in order to tailor mechanical properties for specific applications and to potentially overcome the inherent strength-toughness trade-off.

Amongst stainless steels manufactured via PBF-LB, 17-4PH (also referred to as 1.4542 or X5CrNiCuNb16-4) is one of the most extensively studied alloys [6]. The microstructural evolution from the as-built state to various heat-treated conditions, including solution-treated, solution-treated and aged, and directly aged states, has been broadly investigated to gain a better understanding of the influence of heat treatments on microstructure [7, 8, 9, 10, 11, 12]. Furthermore, its mechanical behaviour under tensile loading has been widely reported in the literature. Compared to conventionally processed counterparts, the PBF-LB 17-4PH has exhibited equivalent or even superior properties in terms of tensile strength and hardness [11, 13, 14, 15]. Huang *et al.* estimated the contributions of various strengthening mechanisms to yield strength (YS) and identified precipitation strengthening as the dominant factor, followed by dislocation, grain boundary, and austenitic phase contributions. Comparing the AM 17-4PH stainless steel to conventional counterparts, the author noted similar dislocation densities and precipitation states, suggesting that the superior tensile properties of the PBF-LB material arise from its finer microstructure [13]. Li *et al.* investigated the effects of post-processing heat treatments on the microstructure and tensile strength of PBF-LB 17-4PH. Ageing treatments were found to increase strength due to the formation of ε-Cu nano-precipitates that impede dislocation motion, while the reduced plasticity was likely due to carbide formation [16]. While the microstructure and tensile properties of 17-4PH have been widely studied, its behaviour under dynamic loading, such as high strain rate conditions, remains underexplored. Kovacs *et al.* examined the impact toughness of PBF-LB 17-4PH, reporting absorbed energies of 22 J (as-built), 9 J (solution-treated), and 6 J (480 °C/1 h - H900) [17]. The low values were mainly attributed to significant lack-of-fusion porosity, without discussing other microstructural factors.

Impact toughness has been gaining increasing attention within the context of PBF-LB, where alloys such as 316L, AlSi10Mg, AlMgScZr, Ni-Cr-Mo based alloy and martensitic steels, amongst others, have been extensively investigated concerning their energy absorption capacity under high strain rate loading [18, 19, 20, 21, 22, 23, 24, 25]. This interest arises from the observed discrepancy in impact toughness values between AM-produced materials and those manufactured by conventional routes, highlighting the potential for performance improvements in AM materials [22, 26, 27]. In particular, significant strength-impact toughness trade-off was demonstrated, while the related mechanisms weren't systematically investigated.

In the category of steels, for example, it has been reported in 316L stainless steel that the failure mechanism and fracture mode transition from transgranular to intergranular is linked with the presence of nanosized Si-rich oxides. This is attributed to heat treatment-induced grain boundary migration, which promotes boundary pinning by oxide particles. This phenomenon leads to a high-volume fraction of oxides at grain boundaries, consequently decreasing impact toughness [18]. On the other hand, several studies have emphasized the significant contribution of strain hardening capability to understanding material behaviour under high strain rate loading [21, 27, 28]. It was claimed that higher strain hardening ability reduces stress concentrations, thereby influencing crack propagation rates [29]. Additionally, within the AM context, the high dislocation densities typically observed in printed components, combined with the high strain rate conditions of testing, may result in a scenario of reduced strain hardening capability [30]. Taking these aspects in account, post-processing heat treatments would be a possible pathway to tailor the as-built AM microstructure in order to break the trade-off between mechanical properties by in particular enhancing impact toughness of PBF-LB manufactured alloys.

Given the limited availability of data on the impact toughness of PBF-LB 17-4PH stainless steel under different post-processing conditions, and considering the broader context of high strain rate loading of AM-fabricated metallic materials, which suggests potential impact toughness enhancements, the need for further studies in this area is evident. This study aims to elucidate the effects of ageing heat treatments on the microstructure and mechanical properties (tensile strength and impact toughness) of PBF-LB 17-4PH, with a particular focus on identifying the deformation and fracture mechanisms under dynamic loading conditions.

## 2. Materials & Methods

*2.1 Sample preparation*

The feedstock powder, supplied by Erasteel, has the chemical composition listed in Table 1. Chemical analysis was performed using optical emission spectrometry (OES), while the nitrogen content was determined by inert gas fusion (IGF) combined with a thermal conductivity detector (TCD). The material was fabricated using an EOS M290 PBF-LB system under a protective argon atmosphere throughout the process. Specimens were produced in the form of square blocks (12 × 12 × 60 mm) and cylinders (12 mm diameter and 60 mm height). The optimized processing parameters included a layer thickness of 40 μm, laser power of 200 W, hatch spacing of 0.1 mm, and a laser scan rotation of 67°. These conditions correspond to a volumetric energy density of 61 J/mm³. The building direction (BD) was aligned parallel to the z-axis, as illustrated in Fig. 1c.

**Table 1:** Chemical composition (wt. %) of the 17-4PH powder and as-built printed sample

| Material | Cr | Ni | Cu | Mn | Si | Nb | N | Mo | S | P | O | C | Fe |
|---|---|---|---|---|---|---|---|---|---|---|---|---|---|
| Standard [31] | 15.00-17.00 | 3.00-5.00 | 3.00-5.00 | 1.00 | 1.00 | 0.15-0.45 | | | 0.030 | 0.040 | | 0.07 | Bal. |
| Powder | 16.39 | 4.11 | 3.75 | 0.34 | 0.71 | 0.28 | 0.03 | 0.015 | 0.006 | 0.008 | 0.047 | 0.035 | Bal. |
| As-built | 16.16 | 4.08 | 3.58 | 0.32 | 0.73 | 0.29 | 0.035 | 0.020 | 0.006 | 0.006 | 0.031 | 0.028 | Bal. |

The as-printed blocks were subjected to heat treatments to assess the effect of ageing on mechanical properties. The thermal cycles included a solution treatment followed by ageing steps, in accordance with established procedures for conventional 17-4PH stainless steel [31]. This involved an initial solution treatment at 1040 °C for 30 minutes, followed by rapid cooling under a nitrogen ($N_2$) atmosphere. Ageing treatments were then applied as follows: 620 °C for 4 h, 620 °C for 8 h and a double-step ageing at 760 °C for 2 h followed by 620 °C for 4 h. The treatment at 620 °C for 4 h corresponds to the standard H1150 condition, while the double-step treatment represents the H1150-M condition, both recommended by ASTM standards for 17-4PH stainless steel [31]. A schematic of the heat treatment procedures employed in this study is presented in Fig. 1a.

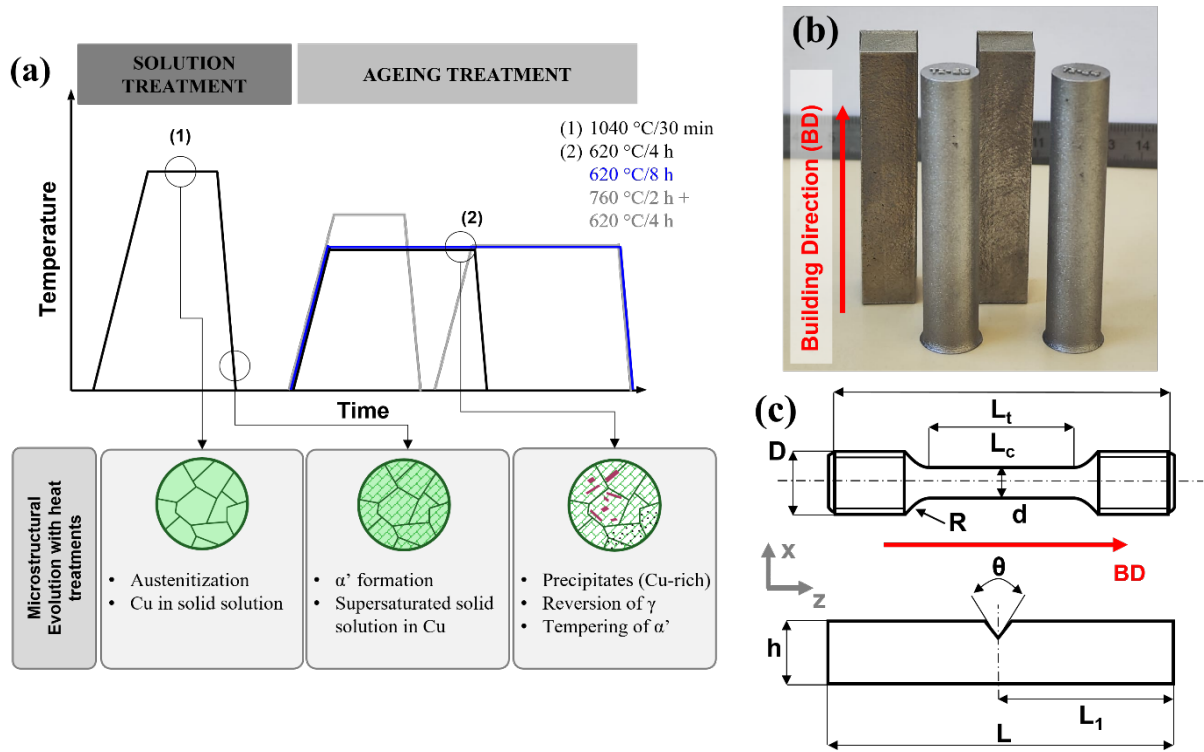

**Figure 1:** Schematic of the conditions and samples used in this study. (a) the studied heat treatments and their influence on the microstructure, (b) photograph of as-built rectangular- and cylinder-shaped blocks along the BD, (c) scheme of the tensile test and Charpy V-notch specimens regarding the BD

*2.2 Microstructural Characterization*

Microstructural characterization was carried out on the zx plane, parallel to the BD of the specimens. The undeformed "reference" region was selected far away from the notch, at the bottom of the half-broken Charpy V-notch specimen, to understand the microstructure prior to deformation, assuming that the stress and strain field were concentrated near the crack tip. To investigate the deformed zones, the Charpy specimens were sectioned in half, and both the area adjacent to the notch and the crack propagation path were examined to assess microstructural evolution under deformation.

For microstructural observations, the samples were prepared following the conventional metallographic steps (grinding and polishing to 1 µm) followed by etching with Fry's reagent (5 g $CuCl_2$, 40 mL HCl, 30 mL, $H_2O$ and 25 mL ethanol) for standard light optical microscopy (LOM) and scanning electron microscopy (SEM) investigations. The etching highlighted the reversed austenite grains on the microstructure and was used to observe the different austenite morphologies as well as the deformed state near the main crack. A vibratory polishing step with colloidal silica suspension was performed on the polished samples in order to reveal the crystallographic contrast for electron backscatter diffraction (EBSD) analysis and image observations.

A Rigaku Smartlab X-ray diffraction (XRD) system equipped with Cu kα radiation source ($\lambda$ = 1.54056 Å) was used for phase analysis. The 2θ angle range was of 35 - 105°, with a step size of 0.01° and step time of 40 s. The volume fraction of austenite was calculated considering the diffraction intensities of body-centred cubic (BCC) and face-centred cubic (FCC) phases as reported in the literature. Since the $111_\gamma$ and $110_\alpha$ diffraction peaks are very close and somehow superposed at their base, in order to effectively estimate the austenite volume fraction, the $200_\alpha$, $211_\alpha$, $220_\gamma$ and $311_\gamma$ peak intensities were taken into account [32].

A Hitachi SU-5000 FEG scanning electron microscope (SEM) equipped with an Oxford instruments symmetry system was used to conduct EBSD analyses, electron channelling contrast imaging (ECCI), energy dispersive spectroscopy (EDS) and backscatter electrons (BSE) and secondary electrons (SE) imaging. EBSD data was acquired with a step size of 0.25 µm for large-area maps, which were employed to characterize the martensitic structure, including prior austenite grain reconstruction and block width measurements. At least 10000 blocks, 3500 packets and 2000 parent austenite grains (PAGs) were taken in account for the martensitic structure measurements. The misorientation threshold used for the identification of blocks was of $\theta > 55°$, meanwhile packet and PAGs were reconstructed after the EBSD data using the MTEX toolbox (version 6.0.0) and ORTools [33]. Since the martensitic microstructure was found to be very heterogenous, a weighted grain size measurement was used for parent grains, packet and block size assessment. A smaller step size of 0.05 μm was used for localized small-area maps, combined with the "Refined Accuracy" mode of band indexing provided by Aztec software. This leads to satisfactory indexing of the nanometric grains and ensure higher angular resolution (0.05°) for kernel average misorientation (KAM) determination. Such a combination allows to take into account the lattice distortion induced by dislocations [34]. At least 1500 grains of reversed austenite were taken in account for the austenitic grain size measurements. EBSD data was treated using HKL Channel 5 software.

Transmission electron microscopy (TEM) was performed using a Thermofisher Titan Themis instrument operated at 300 kV, and equipped with a probe aberration corrector and a highly efficient (4 quadrants) EDS system. The feedstock powder sample for TEM was directly prepared by focus ion beam (FIB) cutting using a Zeiss Crossbeam 550L dual beam microscopy. The PBF-LB samples were mechanically ground and thinned down to around 80 μm, followed by twin jet electro-polishing performed using 10 % perchloric acid and 90 % ethanol with a voltage of 30 V at -30 °C [35]. TEM bright-field imaging was used for statistical martensite lath size measurements. Identification of nanoscale microstructure features (e.g, precipitate, inclusion and powder particle surface skin) was done using high-angle annular dark field (HAADF) imaging associated with EDS mapping in scanning TEM (STEM) mode.

*2.3 Mechanical Characterization*

Tensile and Charpy V-notch specimens were machined from the printed and heat-treated blocks and cylinders to ensure a surface finish compliant with the relevant testing standards. As shown in Fig. 1c, the tensile specimens had a dog-bone geometry with a gauge length (Lc) of 33 mm and a diameter (d) of 6 mm, with the loading direction parallel to the BD. Tensile tests were performed at the constant strain rate of $8 \times 10^{-3}$ s$^{-1}$ and conducted at room temperature using a 3R Syntech 300S testing machine. A mechanical extensometer was used to measure strain in accordance with ISO 6892-1 [36].

Charpy V-notch specimens with a square cross-section of 10 mm (h) and dimensions of 55 mm (L), 27.5 mm (L1), and a notch angle of 45° (θ) were machined in accordance with ISO 148-1 standards [37]. The tests at room temperature were conducted on a PSD 300/150 machine with pendulum working capacity of 300 J. The loading direction was oriented perpendicular to the build direction (BD), as illustrated in Fig. 1c. Two tensile and Charpy specimens were tested for each as-built and heat-treated conditions.

## 3. Results

### 3.1. Microstructure of the feedstock powder

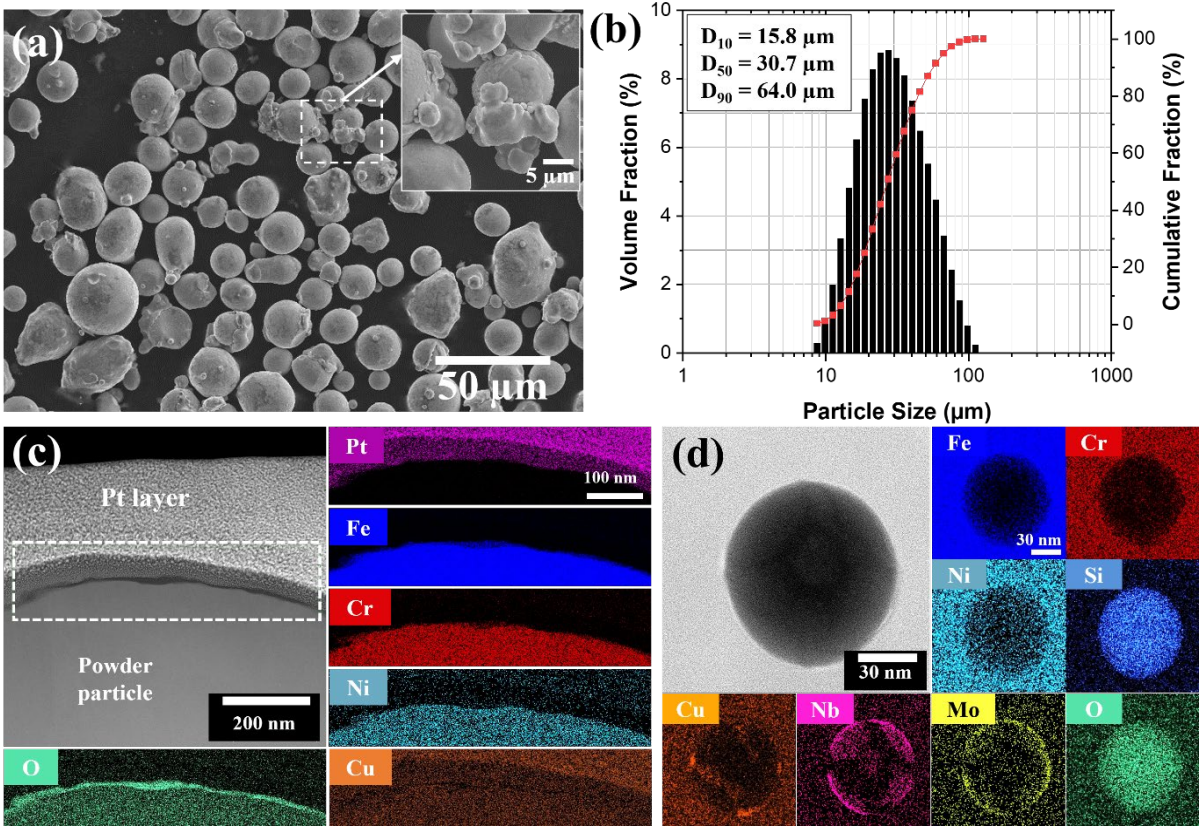

**Figure 2:** The 17-4PH powder used in this study: (a) SEM image showing particle morphology, (b) statistical particle size distribution measured by laser granulometry, (c) HAADF STEM image of an individual particle sample cut by FIB with corresponding EDS elemental maps taken from the white dashed box, note that Pt layer originates from FIB cutting process and (d) HAADF STEM image with corresponding EDS elemental maps highlighting an oxide inclusion inside the analysed particle

The feedstock powder particles exhibit an overall spherical morphology, as shown in Fig. 2a. Small satellite particles are visible on the surfaces of larger particles. A few particles display irregular shapes and appear agglomerated, suggesting partial melting commonly associated with recycled powders. The particle size distribution is nearly Gaussian, with $D_{10}$, $D_{50}$, and $D_{90}$ values of 15.8 µm, 30.7 µm, and 64.0 µm, respectively (Fig. 2b). The distribution indicates good homogeneity in particle size, as evidenced by the narrow histogram concentrated primarily within the 15-45 µm range. Fig. 2c reveals a continuous nanometric oxide layer, with a thickness below 25 nm, uniformly covering the particle surface. No discrete surface oxide particles were observed. Internally, no segregation of the principal alloying elements of Cr, Ni, and Cu was detected within the Fe-rich matrix. However, several nanoscale spherical particles (70-110 nm in diameter) exhibiting a darker contrast were observed. They were enriched in Si and O, with surface segregation of Cu, Nb, and Mo (one example shown Fig. 2d). Consistent with the result reported in reference [38], these observations suggest that the powder is susceptible to oxidation and contamination during the atomization process and/or storage. Moreover, they indicate the non-negligible formation of Si- and O-rich inclusions within the feedstock powder.

*3.2. Microstructure of the PBF-LB as-built sample*

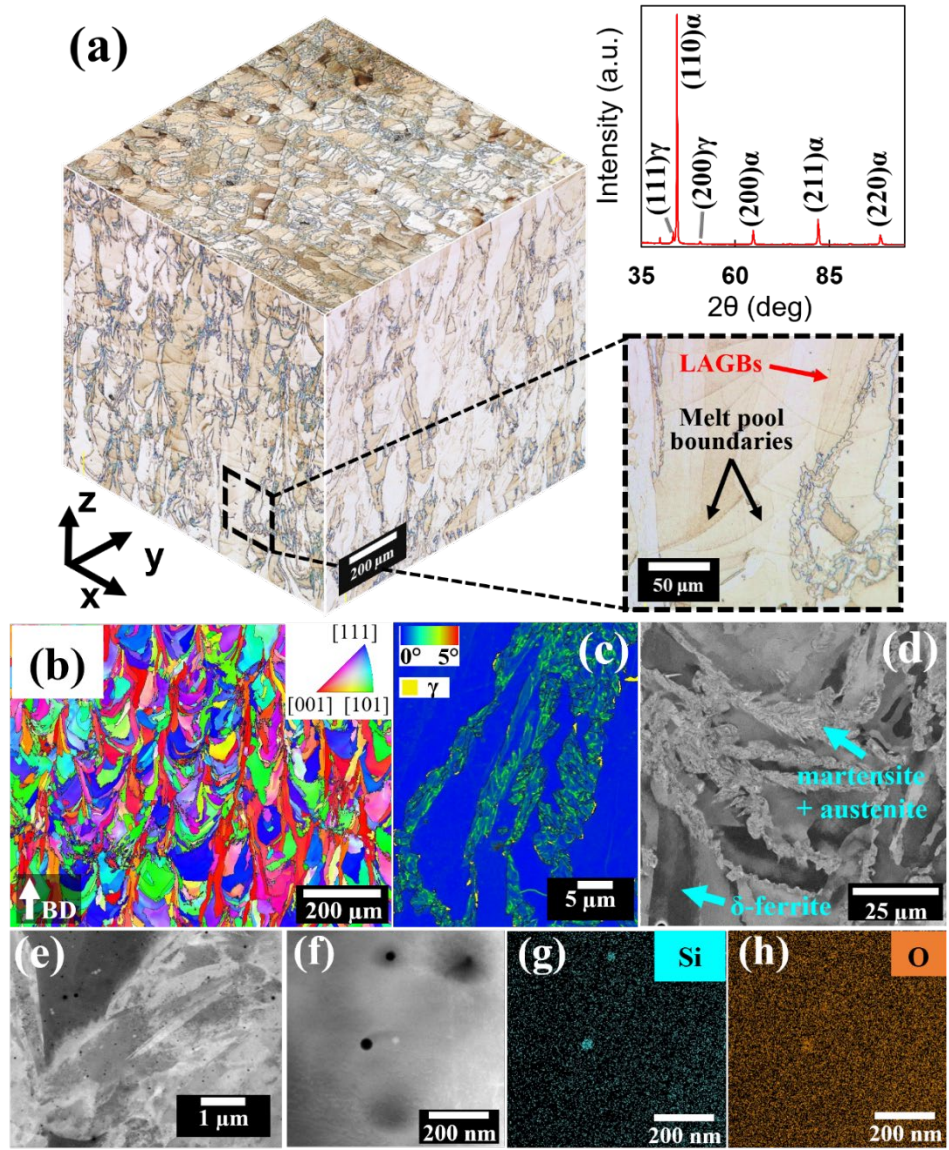

**Figure 3:** Microstructure of the as-built 17-4PH sample: (a) reconstructed 3D LOM micrographs and XRD pattern, (b) orientation map along the z direction (BD), (c) KAM map, (d, e) BSE images with crystallographic contrast, (f) HAADF STEM image, and (g, h) corresponding elemental EDX maps

The microstructure of the as-built sample is presented in Fig. 3. The XRD pattern, acquired on the plane parallel to the build direction (in the zx plane), is shown in Fig. 3a, where the characteristic $110_\alpha$, $200_\alpha$, $211_\alpha$, and $220_\alpha$ reflections of the BCC α phase are clearly identified. However, within this α-BCC matrix, the exact nature of the phase, primarily martensite or δ-ferrite, remains a subject of discussion in the literature, as it can depend on the specific alloy composition; therefore, further microstructural analysis was necessary to conclusively determine the phase constituents [12, 39]. In addition, the $111_\gamma$ and $200_\gamma$ peaks associated with the FCC austenite phase are also detected, indicating the presence of retained austenite. Quantitative phase analysis based on peak intensity ratios estimated the austenite volume fraction to be approximately 5.3 % in the as-built microstructure.

Taking a closer look into the morphology of the studied structure, also evidenced by the Fig. 3a, the plan parallel to the BD presents coarse grains elongated in this direction with the size of dozens of micrometers. Fine grains of a few micrometers and needle-like morphology are found to surround the coarse elongated grains. In Fig. 3a, a section of the microstructure is highlighted, where it is possible to observe some melt pool boundaries, which are typical of the as-built microstructure from the PBF-LB process. The melt pool sizes varied from 20 to 80 µm, being coherent with the processing parameters used in this work. Inside the coarse elongated grains, the contrasts of low angle grain boundaries (LAGBs) are evidenced and slightly etched.

The orientation map parallel to the BD is displayed in Fig. 3b, showing the as-built microstructure does not present preferential orientation of the grains. The KAM map from Fig. 3c shows misorientations of around 2° inside the coarse grains, which correspond to the low-contrast LAGBs observed by LOM. Also, nanometric yellow grains are highlighted on the edges of the fine grain region, and have been indexed as the austenite phase. Furthermore, it is possible to associate the fine-grain regions with low misorientations (2° to 5°) areas, whereas the coarse grains present no misorientations spreads apart the detected LAGBs. The misorientations detected by the KAM maps are associated to the presence of geometrically necessary dislocations (GNDs) and deformation [40], and therefore indicate the fine grains present higher dislocation density than the coarse elongated grains. These observations are in agreement with the literature of the PBF-LB 17-4PH, where the elongated coarse grains have been found to be the δ-ferrite constituent, the fine grains found surrounding the coarse grains correspond to the martensitic constituent and nanometric grains within the martensitic fine grains are the retained austenite [7, 9, 39]. The constituents and morphology assessment would then lead us to conclude the powder composition used in this study generated an as-built microstructure which is composed mainly of δ-ferrite (≈ 70 %), with considerable fractions of martensite (≈ 25 %) and small fractions of retained austenite (≈ 5 %), and Fig. 3d evidences further these constituents' distribution.

Besides the constituents that compose the material's matrix, nanometric particles with dark contrast were detected in the as-built microstructure, as displayed in Fig. 3e. In the image, a bright area corresponding to the martensitic grain region is spotted, underneath a grey zone which was identified as a δ-ferrite grain. The black particles are found evenly spread all over both regions, showing no particular distribution on boundaries or specific constituents. The results displayed in Fig. 3 are typical for the as-built microstructure. The average particle size measured was of 44 ± 22 nm. To further investigate the nature of the observed particles, chemical analysis was conducted using STEM-EDS. The elemental maps shown in Fig. 3f confirm that these dark particles are $SiO_2$ inclusions. As noted in the previous section, nanometric $SiO_2$ inclusions were initially present in the 17-4PH feedstock powder, and it is likely that they remained stable throughout the printing process. However, a quantitative assessment of the evolution of the $SiO_2$ population is beyond the scope of this study.

*3.3. Microstructure of the PBF-LB heat-treated samples*

XRD analysis was conducted on the heat-treated samples to identify the phases with BCC and FCC crystal structures and to estimate their respective phase fractions. The low carbon content of the studied material, resulting in marginal distortion of the C-axis [12], along with the heat treatment routes involving austenitization and quenching, indicates that the BCC-structured phase can be attributed to martensite. The FCC-structured phase corresponds to reversed austenite, which forms during the ageing step as a result of diffusion and partitioning of austenite-stabilizing elements. Fig. 4 presents the XRD patterns corresponding to the 620 °C/4 h, 620 °C/8 h, and 760 °C/2 h + 620 °C/4 h heat treatment conditions. Peaks associated with the BCC phase are clearly visible in all the cases. The intensity of the $(111)\gamma$ reflection, indicative of the FCC austenite phase, is the lowest for the 620 °C/4 h condition and increases with both extended ageing time (620 °C/8 h) and the double-step ageing treatment (760 °C/2 h + 620 °C/4 h). Quantitative analysis based on the intensity ratio of the phases indicates that the 620 °C/4 h microstructure contains approximately 6.8 % austenite, while the 620 °C/8 h and 760 °C/2 h + 620 °C/4 h conditions exhibit approximately 15.2 % and 20.1 % austenite, respectively. These results suggest that incorporating a higher temperature ageing step at 760 °C for 2 h is more effective at promoting austenite reversion than simply extending the ageing duration at 620 °C, despite the shorter total treatment time compared to the 620 °C/8 h condition.

SEM analysis was employed to further investigate the morphology of the reversed austenite. The applied etching procedure selectively revealed the austenitic phase, exposing two distinct morphologies present in the heat-treated microstructures: equiaxed and elongated austenite. The equiaxed grains, which appear coarser, are predominantly located along prior austenite grain (PAG) boundaries, while the finer, elongated austenitic grains are distributed within and between martensitic blocks. Both morphologies are depicted in Fig. 4b-d, where the red arrows indicate the equiaxed grains, and the yellow arrows indicate the elongated morphology. In the 760 °C/2 h + 620 °C/4 h condition, the austenite fraction is sufficiently high that distinguishing between the two morphologies becomes challenging. Similar observations of reversed austenite morphologies in PBF-LB 17-4PH stainless steel have been reported by Huang *et al.* [13]. The austenite grain size was evaluated using EBSD analysis. Despite the 620 °C/4 h condition exhibiting the lowest austenite fraction (6.8%), it presented the largest average austenite grain size of 334 nm. As shown in Table 2, the austenite fraction increases while the average grain size decreases. This trend can be attributed to the progressive development of the finer, elongated austenitic morphology located within and between martensitic blocks, which becomes more prominent in the 620 °C/8 h and 760 °C/2 h + 620 °C/4 h conditions compared to the 620 °C/4 h condition. Since the grain size associated to this morphology is finer, the mean austenitic grain size decreases with the heat treatments.

In addition, the concentration of austenite-stabilizing elements was measured for the 620 °C/4 h and 760 °C/2 h + 620 °C/4 h conditions, which correspond to the lowest and highest austenite fractions, respectively, and also exhibit distinct morphology distributions. The elemental compositions are shown in Fig. 4b and d. For the 620 °C/4 h condition, both equiaxed and elongated austenite morphologies displayed similar concentrations of stabilizing elements; therefore, the overall austenite composition was considered representative for comparison between the two conditions. Despite some experimental scatter, the 620 °C/4 h microstructure exhibited higher Cu and Ni contents compared to the 760 °C/2 h + 620 °C/4 h condition.

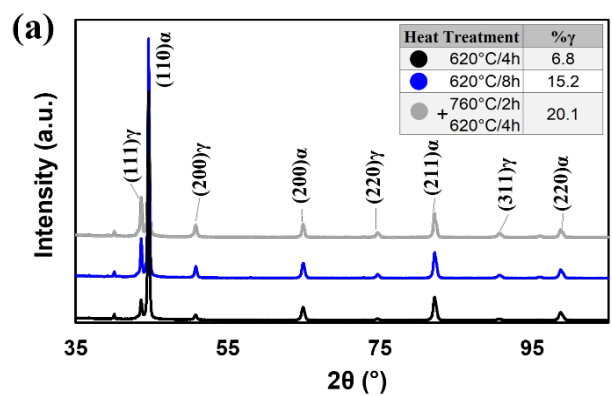

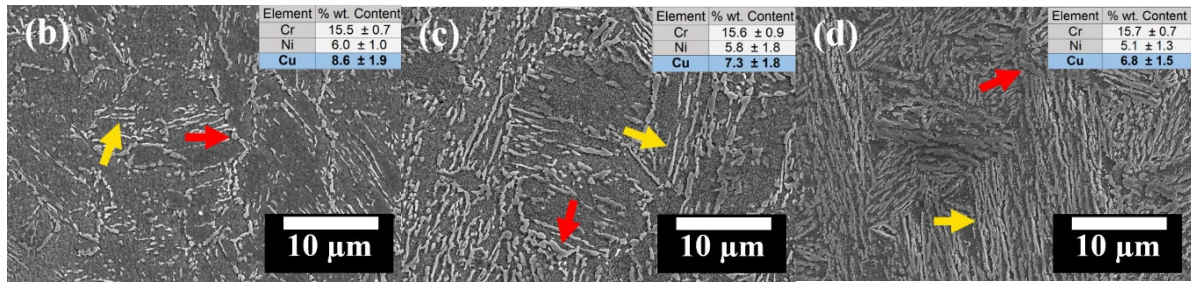

**Figure 4:** Characterization of the reversed austenite: (a) XRD patterns of the heat-treated samples, and SEM-SE images of the austenite morphologies with the quantification of the austenite-stabilizing elements for (b) 620 °C/4 h, (c) 620 °C/8 h and (d) 760 °C/2 h + 620 °C/4 h conditions

The first stage of the heat treatments investigated in this study is the solution treatment, designed to produce a copper-supersaturated solid solution for subsequent precipitation and to austenitize the microstructure. To assess the microstructural evolution throughout the heat treatment sequence, PAGs were reconstructed from EBSD data acquired on the as-quenched condition, retracing the austenitic phase without considering the tempering effects introduced during the ageing step. Since all heat-treated samples underwent the same solution treatment (1040 °C for 30 min), identical austenitization conditions were applied, and thus the PAG structure should be identical across all conditions. The reconstruction was successful for approximately 90 % of the data, as shown in Fig. 5a. The reconstructed austenitic structure appears heterogeneous, comprising both coarse grains elongated along the BD and very fine grains. This suggests that the solution treatment at 1040 °C for 30 minutes was insufficient to fully recrystallize the microstructure and eliminate the thermomechanical history imparted by the PBF-LB process. Nevertheless, although the grain structure remains heterogeneous, the characteristic melt pool traces visible in the as-built condition (Fig. 3b) are no longer evident in the reconstructed PAGs. The average PAG size was measured to be 23.2 μm.

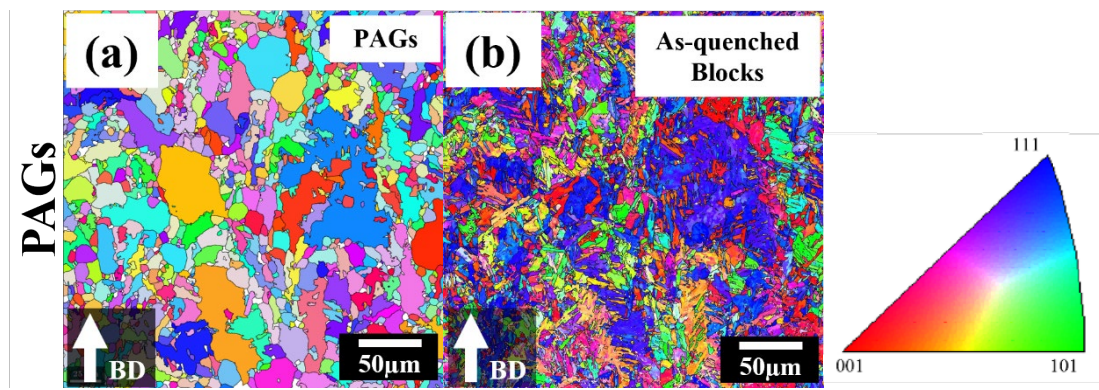

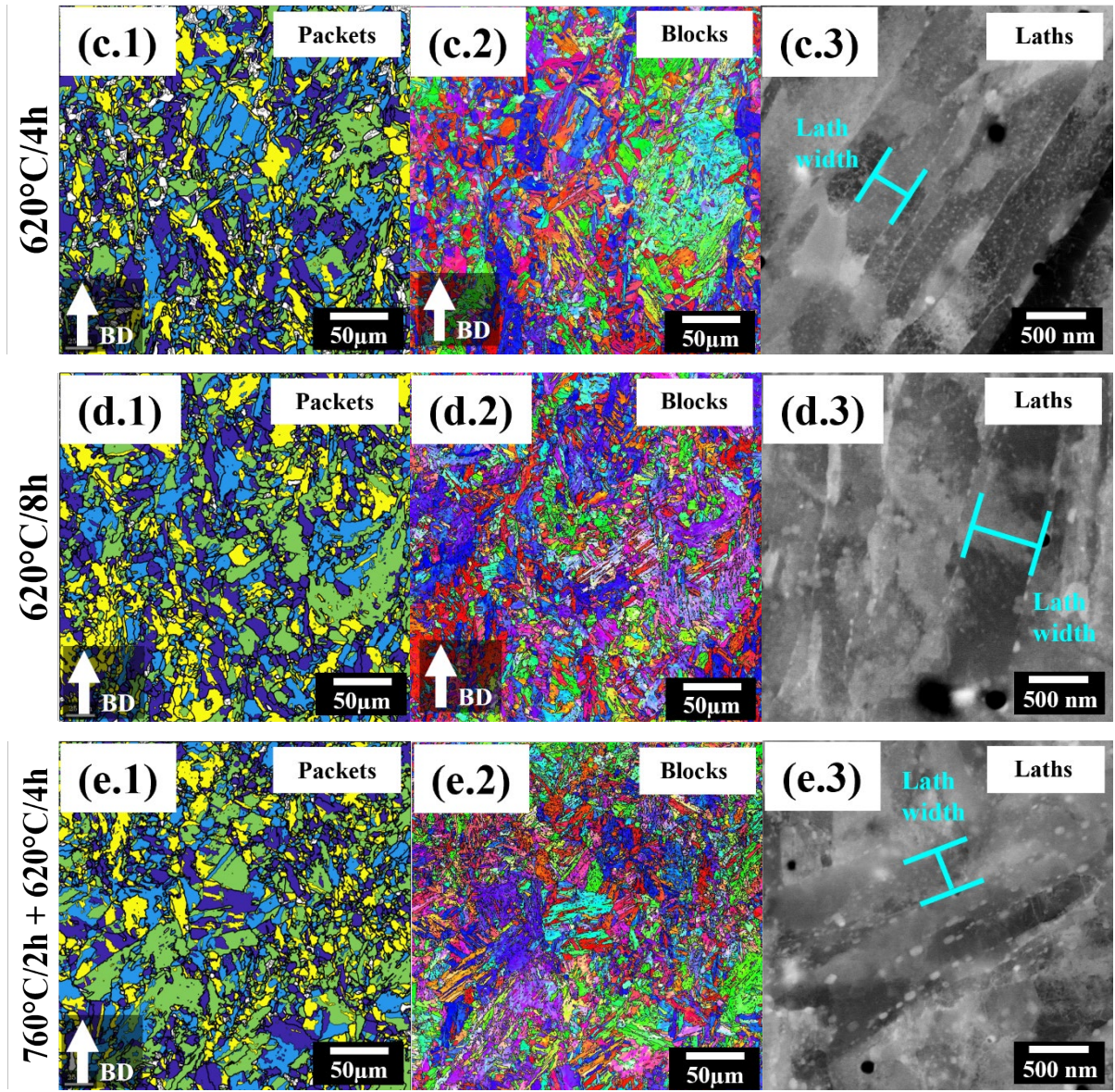

**Figure 5:** Microstructure evolution of PBF-LB 17-4PH with the heat treatments: (a) Reconstructed PAG's mean orientation, (b) EBSD orientation map of the as-quenched structure used for the PAG reconstruction, (c.1 – e.1) reconstructed packets, (c.2 – e.2) EBSD orientation map evidencing blocks, (c.3 – e.3) ECCI evidencing the martensitic laths. The orientation maps and PAG's mean orientation are taken regarding the direction parallel to BD

Following the solution treatment step, the material underwent ageing at various temperatures and durations, promoting austenite reversion, martensite tempering, and the precipitation of Cu-rich particles. A comparison of the microstructures for each heat treatment condition, as shown in Fig. 5, reveals no significant qualitative differences between the aged structures. To assess the influence of martensite tempering during the ageing process, the hierarchical martensitic structure was quantitatively analysed. In a first attempt, the ECCI investigations (see Fig. 5c.3, 5d.3, 5e.3) indicate that the tempered martensitic structure remains rather similar for all the studied conditions. Here, only a slight coarsening of the lath is observed for the longest ageing time (8h). Table 2 summarizes the measured values for packet size, block width, and lath thickness across the different heat-treated conditions. The packet size increased from the 620 °C/4 h to the 620 °C/8 h condition but decreased when the 760 °C/2 h + 620 °C/4 h treatment was applied, compared to the 620 °C/4 h condition. The largest block width was observed in the 620 °C/4 h microstructure, while smaller block widths were found for both the 620 °C/8 h and 760 °C/2 h + 620 °C/4 h conditions. These trends are consistent with the orientation maps, where the 620 °C/4 h condition visibly exhibits larger blocks compared to the other heat-treated states. Lath thickness increased with prolonged ageing time, as seen in the 620 °C/8 h condition, suggesting that extended heat treatment promotes lath coarsening and coalescence. In contrast, the 620 °C/4 h and 760 °C/2 h + 620 °C/4 h conditions displayed comparable lath sizes.

**Table 2:** Martensitic hierarchic structure and reversed austenite grain sizes

| Heat Treatment | PAG Size (μm) | Packet Size (μm) | Block Width (μm) | Lath Size (nm) | γ Grain Size (nm) |
|---|---|---|---|---|---|
| 620°C/4h |  | 18.8 ± 2.5 | 11. ± 2.1 | 238 ± 110 | 334 ± 3 |
| 620°C/8h | 23.2 ± 0.9 | 20.8 ± 2.8 | 8.9 ± 1.8 | 272 ± 145 | 285 ± 37 |
| 760°C/2h + 620°C/4h |  | 16.2 ± 1.9 | 8.5 ± 1.7 | 232 ± 88 | 273 ± 6 |

SEM-BSE imaging further revealed the presence of nanoscale particles with dark contrast dispersed throughout the matrix of the heat-treated samples, as indicated by the red arrows in Fig. 6a–c. These dark particles were frequently, though not exclusively, located along high-angle grain boundaries (HAGBs), including block and PAG boundaries. To clarify their chemical nature, an EDS line scan was performed, confirming that these particles are enriched in silicon and oxygen, consistent with $SiO_2$ inclusions, as shown in Fig. 6e. As all the three heat-treated conditions exhibited similar chemical signatures, this indicates that $SiO_2$ inclusions persist throughout the entire PBF-LB manufacturing process from the initial powder to the as-built microstructure and through subsequent heat treatments. The $SiO_2$ inclusions were further characterized in terms of particle size (Fig. 6g) and size distribution (Fig. 6h), revealing average sizes of 164 ± 72 nm, 124 ± 59 nm, and 143 ± 78 nm for the 620 °C/4 h, 620 °C/8 h, and 760 °C/2 h + 620 °C/4 h conditions, respectively. Compared to the as-built state (44 ± 22 nm), these results indicate significant coarsening of the inclusions during heat treatment, with particle sizes increasing by a factor of 3 to 4. The particle size distributions for the 620 °C/8 h and 760 °C/2 h + 620 °C/4 h conditions were similar, whereas the distribution for the 620 °C/4 h condition was shifted toward larger sizes, with a greater presence of particles around 150 nm.

The broader size distribution observed in the heat-treated samples relative to the as-built condition suggests a more heterogeneous particle population resulting from the thermal exposure. In addition to these dark inclusions, sporadic white nanoparticles were also observed (highlighted by white arrows in Fig. 6a–c). These particles showed no clear preferential distribution but were occasionally associated with the $SiO_2$ inclusions. An EDS line scan (Fig. 6f) revealed that these white particles were enriched in niobium and depleted in iron and chromium, suggesting the formation of Nb-rich phases.

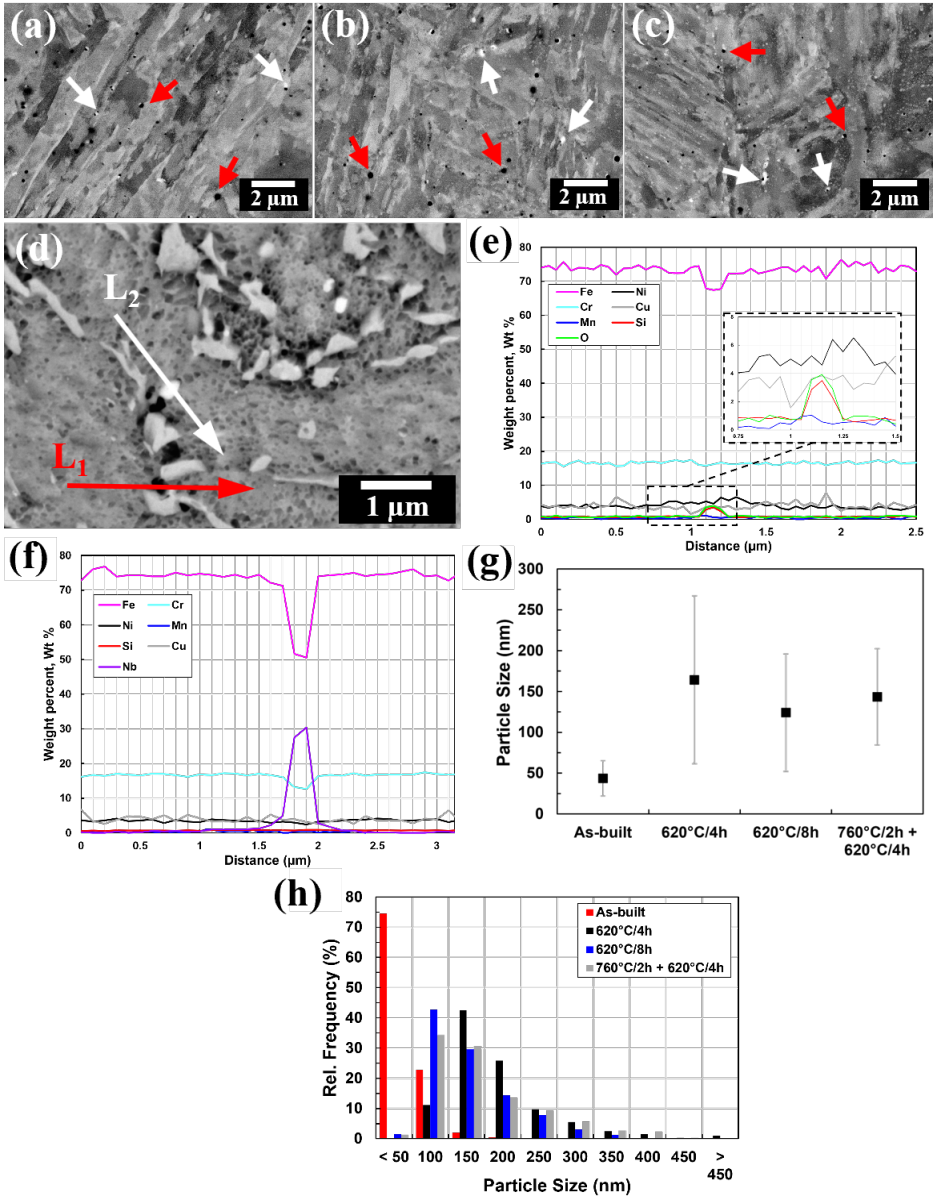

**Figure 6:** SEM-BSE images showing $SiO_2$ and Nb-rich particles in the samples subjected to (a) 620 °C/4 h, (b) 620 °C/8 h, and (c) 760 °C/2 h + 620 °C/4 h treatments. EDS Line-scan performed on the 620 °C/4 h microstructure (d) BSE image with measurements site, (e) a SiO$_2$ particle (black contrast) and (f) a Nb-rich particle (white contrast), (g) average particle size and (h) particle size distribution of SiO$_2$.

Finally, SEM and TEM analyses were employed to investigate the formation of Cu-rich nanoprecipitates, as shown in Fig. 7. These precipitates appear with bright contrast in both SEM-BSE and HAADF STEM images. In the 620 °C/4 h microstructure (Fig. 7a.1–a.3), Cu-rich particles are finely and homogeneously dispersed within the martensitic matrix, with an average particle size of 24 nm. Similarly, the 620 °C/8 h condition (Fig. 7b.1–b.3) exhibits a uniform distribution of fine precipitates, although with a slightly increased average size of 28 nm compared to the 620 °C/4 h treatment. The green circles in Fig. 7a.1 and 7b.1 highlight this homogeneous precipitation along the martensitic laths in both conditions. In contrast, the 760 °C/2 h + 620 °C/4 h microstructure (Fig. 7c.1–c.3) reveals a heterogeneous precipitation state. Areas depleted of precipitates can be identified (indicated by the green arrow in Fig. 7c.1), alongside regions containing homogeneously dispersed fine precipitates (green circle). Additionally, coarser Cu-rich precipitates are observed at the periphery of the precipitate-depleted zones. The average Cu-rich particle size measured for this condition was approximately 30 nm; however, the particle size distribution (Fig. 7e) indicates the presence of two distinct populations: a fine particle population with an average size of 21 nm, and a coarse population averaging 59 nm. The measured precipitate surface fractions were approximately 5.7 % and 5.8 % for the 620 °C/4 h and 620 °C/8 h microstructures, respectively, consistent with their similar precipitation states. In contrast, the 760 °C/2 h + 620 °C/4 h condition exhibited a lower precipitate surface fraction of approximately 3.5 %, reflecting its heterogeneous precipitation state characterized by regions completely depleted of precipitates and the presence of coarser particles.

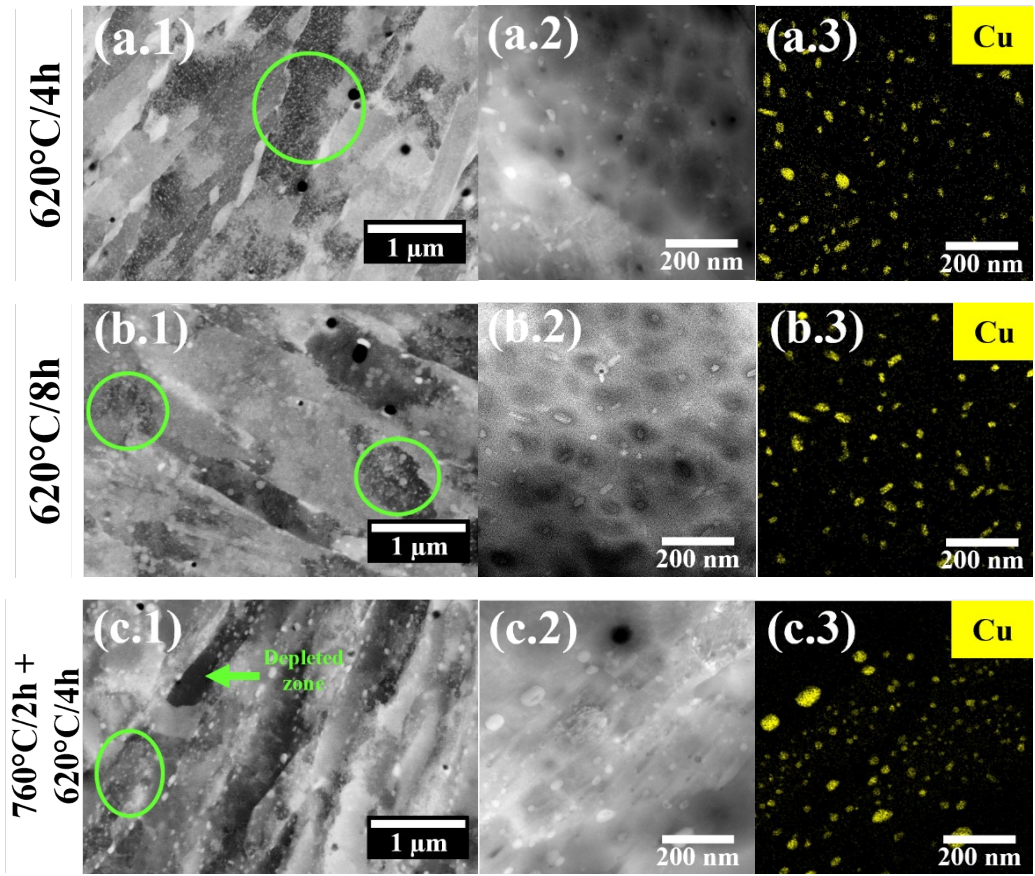

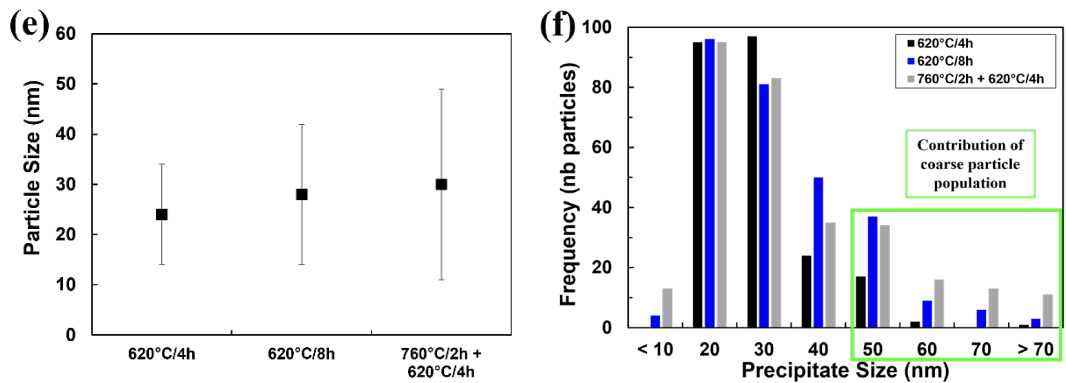

**Figure 7:** Characterization of Cu-rich nanoprecipitates in the matrix: (a1 – c1) SEM-BSE images, (a2 – c2) HAADF STEM images with higher magnification, (a3 – c3) STEM-EDS Cu elemental map, (e) average Cu-rich particle size and (f) corresponding particle size distribution. The conditions represented are (a1 – a3) 620 °C/4 h, (b1 – b3) 620 °C/8 h, and (c1 – c3) 760 °C/2 h + 620 °C/4 h.

*3.4. Mechanical Properties*

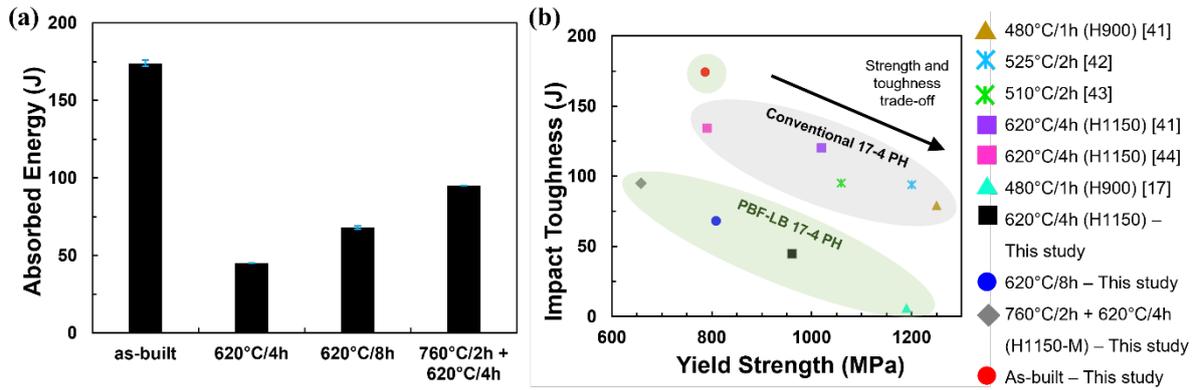

**Figure 8:** Mechanical properties of the as-built and heat-treated 17-4PH steel samples: (a) Impact toughness values and (b) impact toughness vs. YS relation for the studied conditions and reported values in the literature [41, 42, 43, 44, 17]

The impact toughness of the heat-treated samples was evaluated based on the absorbed energy measured during Charpy V-notch tests performed at room temperature, as shown in Fig. 8a. Among all conditions, the as-built state exhibited the highest absorbed energy, reaching an average of 174 J. This superior impact performance may be attributed to its microstructural composition, consisting of approximately 70 % δ-ferrite, a relatively soft phase that enhances toughness. In contrast, the sample subjected to a 620 °C/4 h heat treatment displayed the lowest impact toughness, with an average absorbed energy of 45 J. Extending the ageing time to 8 hours at the same temperature improved the absorbed energy to 68 J. The detailed values of impact toughness measured for each specimen of the studied conditions can be found in Table 3. The highest impact toughness among the heat-treated conditions was obtained with the two-step treatment of 760 °C/2 h followed by 620 °C/4 h, which resulted in an absorbed energy of 95 J. Additionally, the relationship between impact toughness and YS, as determined from tensile tests, was examined for all processing conditions and is presented in Fig. 8b. The results clearly demonstrate the typical trade-off between strength and toughness: higher YS generally corresponds to lower impact toughness, and vice versa. For comparison, relevant data from the literature are included, covering both AM PBF-LB processed and conventionally manufactured 17-4PH steel. When compared to these references, the 620 °C/4 h heat treatment in this study achieves YS values that are similar to or even surpass those of the conventionally produced alloy, indicating that the tensile performance of the AM PBF-LB material is at least maintained. However, this improvement in strength comes at the expense of impact toughness, which remains significantly lower. Specifically, there is an approximate 80 J gap in absorbed energy between the conventionally manufactured 17-4PH steel and the AM PBF-LB sample treated at 620 °C/4 h. A similar reduction in toughness is also reported for the 480 °C/1 h (H900) heat treatment investigated by other authors.

Table 3: Yield strength and impact toughness for each specimen tested.

| Heat Treatment | Reference | Yield Strength (MPa) | Impact Toughness (J) |
|---|---|---|---|
| As-built | specimen 1 | 777 | 175 |
| | specimen 2 | 796 | 172 |
| 620°C/4h | specimen 1 | 956 | 46 |
| | specimen 2 | 965 | 43 |
| 620°C/8h | specimen 1 | 805 | 68 |
| | specimen 2 | 813 | 68 |
| 760°C/2h + 620°C/4h | specimen 1 | 659 | 95 |
| | specimen 2 | 658 | 95 |

Fig. 9 presents the post-mortem fracture surfaces of the Charpy V-notch specimens for each of the heat-treated conditions. Examination of these surfaces provides insight into the fracture mechanisms operating under high strain rate loading. For the specimen heat-treated at 620 °C/4 h, the fracture surface reveals a predominance of dimples, indicative of ductile fracture, along with the presence of planar facets characteristic of cleavage. These planar facets exhibit river patterns that delineate cleavage steps, which likely originate from orientation changes within the microstructure, such as packet, block, and lath boundaries. Cleavage features were primarily observed in the central region of the fracture surface, while the shear lips and final fracture zones contained only dimples, suggesting localized ductile behaviour in these areas. The coexistence of dimples and cleavage facets indicates a mixed fracture mode, involving both micro-void coalescence and cleavage fracture mechanisms. In contrast, the specimens subjected to the 620 °C/8 h and the two-step 760 °C/2 h + 620 °C/4 h heat treatments displayed fracture surfaces covered entirely with dimples, suggesting a fully ductile fracture mode for these conditions. Particles were systematically observed at the bottom of the dimples and were identified as $SiO_2$ inclusions. No evidence of intergranular fracture was detected in any of the conditions.

It should be noted that fracture surface analyses from tensile tests are not included in this study, as the tensile fracture behaviour of PBF-LB 17-4PH steel has already been extensively documented in the literature. While the primary focus of this work is on the critical impact properties, the strength–toughness trade-off and the corresponding deformation and fracture mechanisms will be further discussed in the following section.

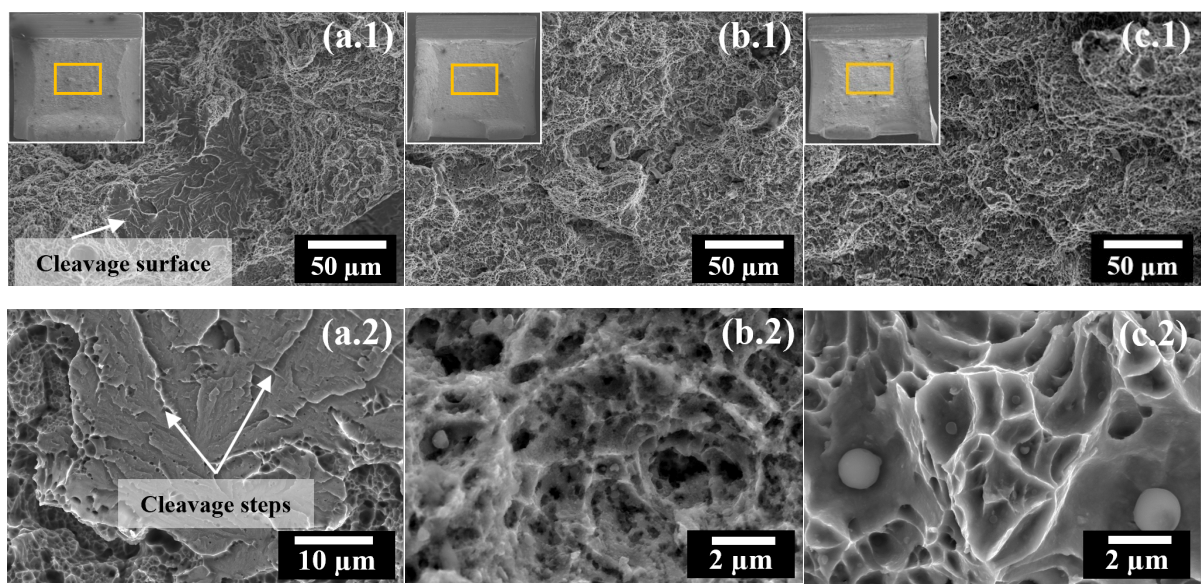

**Figure 9:** SEM-SE images showing fracture surface of the (a1, a2) 620 °C/4 h, (b1, b2) 620 °C/8 h and (c1, c2) 760 °C/2 h + 620 °C/4 h specimens after impact test. Images in (a2 - c2) are magnified from the corresponding box areas as inset in (a1 - c1) taken in the centre of fracture surface

## 4. Discussion

The observed evolution of mechanical properties with different heat treatments prompted a detailed analysis of the corresponding microstructural changes induced by ageing. Three key microstructural factors were identified as influencing both impact toughness and YS: the size and distribution of $SiO_2$ inclusions, the stability of reversed austenite and Cu-rich precipitates, and variations in the chemical composition of the martensitic matrix. In the following discussion, particular attention will be given to the embrittlement promoted by $SiO_2$ inclusions at HAGBs, the enhancement of plasticity associated with the presence of reversed austenite, and the strengthening effect of Cu-rich nanosized precipitates on the martensitic matrix.

### 4.1. The effect of $SiO_2$ inclusions on impact toughness

De Sonis *et al.* investigated PBF-LB 316L stainless steel under impact loading and found that fracture occurred via interfacial decohesion between hard oxide particles and the soft matrix. Heat treatment-induced recrystallization led to particle segregation at grain boundaries, promoting intergranular fracture [19]. Similar findings, as shown in Fig. 7, evidences Si oxide inclusions often found distributed along HAGBs, in particular block boundaries, of the heat-treated samples. Furthermore, micro-cavities were discovered all over the deformed microstructure, specially near the crack (in rectangle area of Fig. 10a). For example, the 620 °C/8 h microstructure is assessed where the lodged particles were found inside these micro-cavities. EDS analysis of these particles indicated they were enriched in Si and O (Fig. 10e), confirming them to be the oxide inclusions previously detected in the undeformed microstructure. All the heat-treated conditions presented similar findings as to the presence of micro-cavities and the elemental profiles of the oxide inclusions lodged inside them. These observations indicate the Si oxide inclusions are the origin of the nucleation of micro-cavities under high strain rate loading. Further investigation into the deformed microstructure revealed micro-cavity coalescence to form micro-cracks on HAGBs (Fig. 10d) which would act in the sense of lowering the crack initiation energy during impact loading [20, 18]. The HAGBs can be slightly seen as the lines that mark a difference in the growth direction of the austenitic phase between neighbour regions, indicated by the white arrows. Hence, the presence of oxide inclusions has a non-neglectable role on the deterioration of impact toughness of the heat-treated specimens, as the heat treatments promoted $SiO_2$ inclusion coarsening and particle distribution on boundaries. Additionally, the powder characterization in Fig. 2 evidenced the presence of a nanoscale oxygen rich layer on the surface of the particles as well as $SiO_2$ nanosized particles inside the powder. These findings would enlighten the origin of these oxide inclusions in the PBF-LB manufactured 17-4PH, considering the presence of $SiO_2$ inclusions is not typical of the microstructure of the conventionally manufactured 17-4PH steel [45, 46, 47].

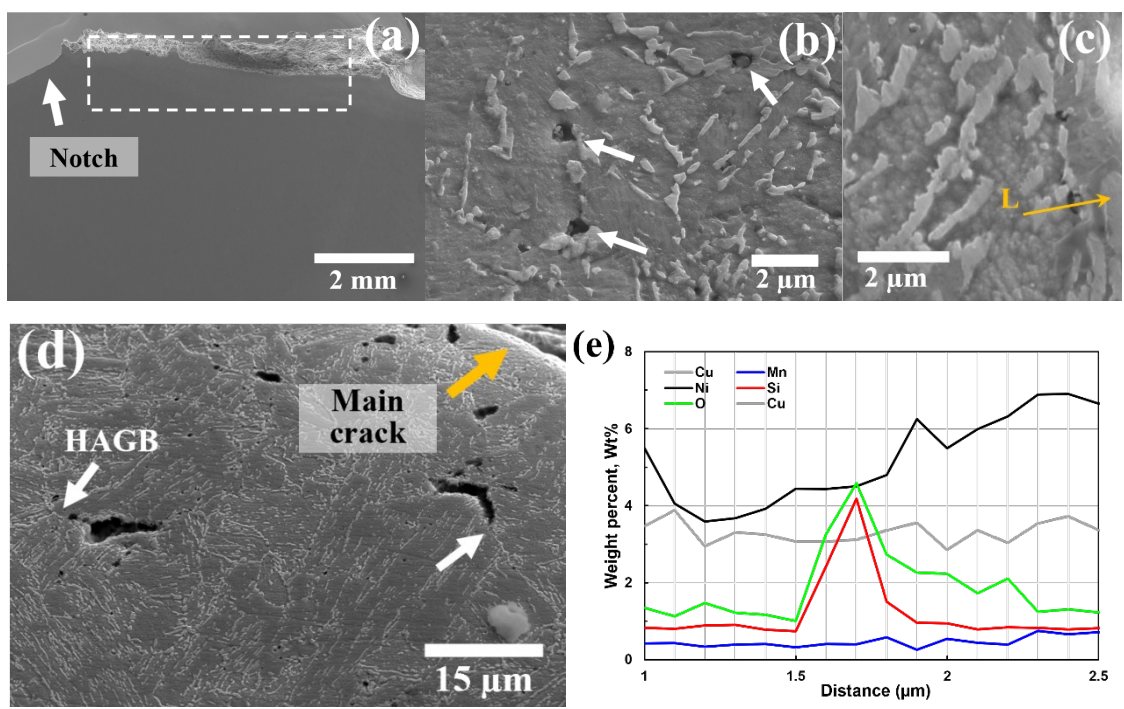

**Figure 10:** SEM analysis of the deformed microstructure of the 620 °C/8 h sample etched with Fry's reagent evidencing (a) the profile of the main crack propagation pathway, (b) micro-cavities found around particles, (c) typical site crossing a particle used for EDS measurement, (d) micro-cavities coalescence on HAGBs and (e) corresponding EDS line scan result along L in (c)

Although our results convey a rather pessimistic message that tailoring the intrinsic microstructural features of 17-4PH steel through (standard and near-standard) ageing heat treatments alone does not appear sufficient to overcome the strength-toughness trade-off (see Fig. 9b), this outcome must be interpreted in the broader context of powder-based manufacturing processes such as AM and powder metallurgy. Unlike conventional wrought or cast processing routes, these methods inherently introduce a significant number of oxide particles in the microstructure. These oxides originate primarily from the initial feedstock powder and may further proliferate due to oxygen uptake during subsequent processing steps. It is therefore widely recognized that minimizing oxide formation through strict control of oxygen content is one of the most effective strategies for achieving a desirable balance between strength and impact toughness. For instance, Cooper *et al.* [48] demonstrated that hot isostatically pressed 316L steel could reach Charpy impact toughness levels comparable to those of its forged counterpart only when the oxygen concentration was tightly controlled between 30 and 60 ppm. A similar trend was reported by Liao *et al.* [49] in the case of fine-grained Mg-3Al-Zn alloys produced by powder metallurgy. While oxygen content had a limited effect on tensile strength, levels exceeding 1000 ppm drastically reduced both ductility and Charpy impact energy, due to impaired plastic deformation and resistance to crack propagation. Consistent conclusions have also been drawn in recent studies on PBF-LB Ni-Cr-Mo based alloys. Peters *et al.* [25] emphasized that controlling oxide inclusion formation is essential for optimizing fracture-related properties in these systems.

*4.2. Transformation-induced plasticity (TRIP) effect and the associated austenite content*

As shown in Fig. 8, impact toughness increased progressively from the 620 °C/4 h to the 620 °C/8 h and 760 °C/2 h + 620 °C/4 h heat treatment conditions. The significantly higher reversed austenite content in the latter condition prior to testing (Fig. 4a) suggests a key role in energy absorption and toughness enhancement. To explore this further, we examined its distribution in highly deformed regions. The deformed microstructure of post-mortem Charpy

V-notch specimens was analysed near the notch, where plastic strain is concentrated [50], and compared to the undeformed state.

Fig. 11 shows coupled image quality (IQ) and phase maps from an undeformed area as well as a zone close to notch for the three studied heat treatment conditions. In a general trend, the austenite content decreases from unstrained to strained areas, confirming the metastable aspect of the austenite that transforms during straining. Nevertheless, some particular features must also be considered: the initial content of austenite, as well as the amount that transforms into martensite greatly differ depending on the heat treatment. Indeed, the more the initial content of austenite (ranging from 6.5 % to 15.3 % for the 620 °C/4 h, 620 °C/8 h and 760 °C/2 h + 620 °C/4 h conditions), the more it transforms (27 %, 80 % and 98 % respectively). Furthermore, the initial morphology of the austenite and its localisation, as well as its change within the accumulated strain also differ depending on the different heat treatment. For the 620 °C/4 h condition, the austenite appears mainly under the form of blocks with an equiaxed shape, decorating the BCC matrix boundaries. Within the deformed region close to the notch, the most part of the austenite follows the deformed shape of the matrix, while some blocks remain unaltered in shape. On the other hand, regarding the two other heat treatments, the austenite still appears in the form of equiaxed grains at the various interfaces of the matrix. However, its shape is mainly elongated. After straining, the matrix is much more deformed in comparison to the first case. The equiaxed austenite blocks tend to follow the change in term of shape of the matrix grains, e.g. accommodation of the plastic strain, the remaining islands have a strong elongated shape, but most of the austenite has been transformed.

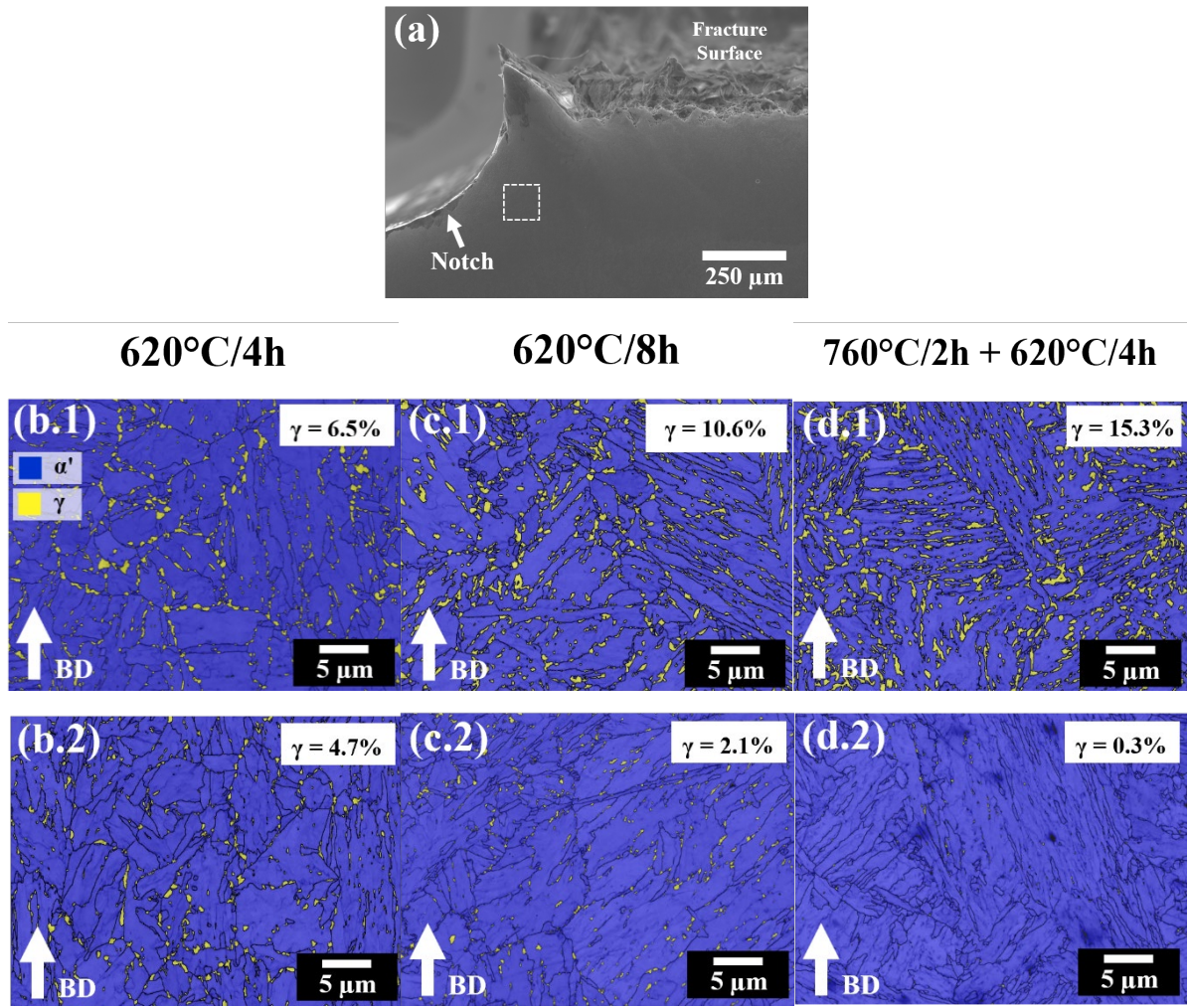

**Figure 11:** Deformed microstructure of the heat-treated conditions showing (a) SEM-SE image indicating the region next to the notch where the EBSD maps were performed, and phase content and distribution measured by EBSD on the (b.1 - d.1) undeformed state and (b.2 – d.2) deformed state. The austenite content in inset is given for each state.

To understand the role played by the reversed austenite, its stability as well as the possible TRIP effect occurring should be considered. Several parameters are known to affect the stability of the austenite. Some of the reported factors for the austenite stabilization under plastic strain are chemical stability, size, morphology, distribution of the austenitic grains and volume fraction [51, 52]. The chemical composition strongly affects not only the metastable austenite state but also the deformation mode of austenite. It seems therefore relevant to refer to metallurgical parameters such as the $M_{d30}$ temperature and the stacking fault energy (SFE). $M_{d30}$ temperature, generally determined by the Equation (1) [53] represents the temperature at which 50 % austenite is transformed into α' martensite at 30 % plastic strain for tensile deformation. Even though the $M_{d30}$ is only a parameter used for quasi-static tensile deformation, here it is used as a mean to comparatively assess the austenite stability under strain.

$$M_{d30} (°C) = 551 - 462(C + N) - 9.2Si - 8.1Mn - 13.7Cr - 29(Ni + Cu) - 18.5Mo - 68\,Nb - 1.42(G - 8) \quad (1)$$

The elements represent their weight fraction on the composition of the austenite phase and *G* represents the austenitic grain size in the ASTM scale. Here Cu and N have a strong influence on the $M_{d30}$ temperature. The content of austenite stabilizing elements was measured on the reference microstructure previous to deformation and the 620 °C/4 h microstructure presented higher Cu (wt %) content than the 760 °C/2 h + 620 °C/4 h one, as displayed in Fig. 4. Cu is a strong austenite phase stabilizer and therefore it confers chemical stability to the austenitic grains by rising the chemical driving force for the martensitic transformation during plastic straining. The origin of the difference of Cu content observed between the microstructures lies on the different ageing steps of the applied heat treatments. The first ageing step of the 760 °C/2 h + 620 °C/4 h heat treatment at 760 °C would promote the precipitation of Cu-rich particles so that the martensitic matrix would be depleted in Cu compared to the 620 °C/4 h which presents only one ageing step, and therefore when the second ageing step is applied, the austenitic grains formed by the reversed martensite into austenite would present a lower Cu partition. The $M_{d30}$ temperature was calculated for the different ageing conditions according to the EDS measurements (see Fig. 4). All the elements were taken into account for the calculation. The $M_{d30}$ values for the 620 °C/4 h, 620 °C/8 h and 760 °C/2 h + 620 °C/4 h conditions are –181 °C, –143 °C and –115 °C, respectively. This is coherent considering the microstructure with the least transformation rate, is the one with the lowest $M_{d30}$ temperature, in this case, the 620 °C/4 h one. Jacob found the similar values of $M_{d30}$ for the PBF-LB 17-4PH stainless steel in his research [54]. Regarding the mechanical stability of austenite by taking into account the SFE and by using the formula given by Meric de Bellefon *et al.* [55], Cu increases the SFE of austenitic matrix [56]. Both lower $M_{d30}$ and higher SFE value result in inhibition of strain-induced martensitic transformation in the 620 °C/4 h steel causing the loss of the TRIP effect.

The deformation accommodated by the martensitic matrix may also be at the origin of the different degrees of the strain-induced transformation. The 620 °C/4 h microstructure exhibits a low level of martensite deformation, as the morphology of the blocks is minimally affected by the plastic strain. This would mean the austenite grains are restrained by the surrounding matrix, which does not undergo significant plastic flow, and therefore cannot accommodate the change in volume associated to the plasticity assisted martensite transformation. Together with the degree of austenite stability depicted by $M_{d30}$ and the SFE, the stress required to trigger the transformation of the austenite in the 620 °C/4 h sample needs to be high so that the stress intensity factor at the SiO$_2$ oxide roots allows crack initiation and propagation in the rather hard, poorly ductile and to some extent brittle martensitic matrix.

As to the influence of the grain size on the austenite stability, coarser grains are less stable under plastic strain, since they present more potential nucleation sites for the martensitic formation [57]. However, this aspect seems to be less important on the microstructural evolution, since the 620 °C/4 h microstructure presented coarser austenitic grains and also the smaller degree of strain-induced transformation.

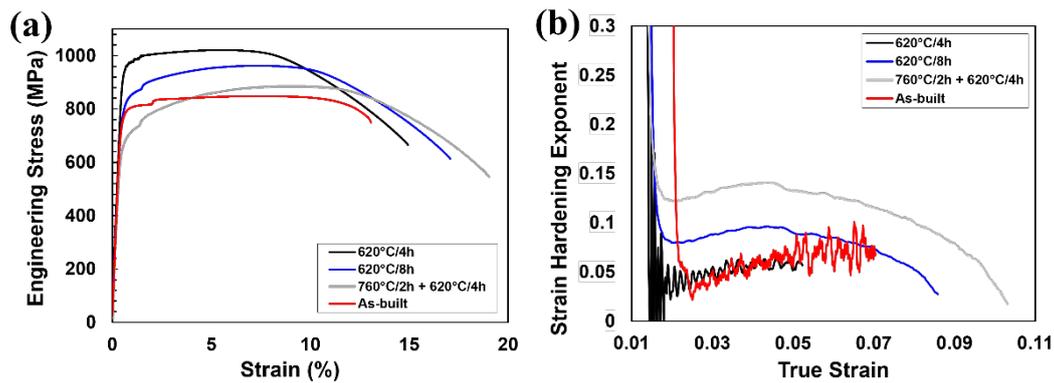

**Figure 12:** Tensile properties of the as-built and heat-treated conditions (a) engineering stress x strain curves and (b) strain hardening curves.

The behaviour under quasi-static tensile loading of the studied materials (Fig. 12) can be mainly related to their precipitation state as well as the TRIP effect capability of each microstructure. The 620 °C/4 h condition exhibited the highest YS (961 ± 5 MPa) and UTS (1022 ± 2 MPa) but the lowest uniform elongation (4.7%) among heat-treated samples. Conversely, the 760 °C/2 h + 620 °C/4 h treatment showed the lowest YS (659 ± 1 MPa) and UTS (885 ± 1 MPa) but the highest uniform elongation (8.6%) (Fig. 12a). The 760 °C/2 h + 620 °C/4 h condition exhibited the highest values of strain hardening exponent compared to the other microstructures, as well as the occurrence of strain hardening throughout a wider range of strain. This is in agreement with the uniform elongation, which is also the highest of the studied conditions, and it translates the fact that the microstructure is able to deform further before the onset of necking conditions under tensile loading [58]. Even though the TRIP effect is responsible for improving elongation, in this study it doesn't seem to act in the sense of increasing significantly the tensile strength values, notably the ultimate tensile strength (UTS). This can be reasonably justified by the different precipitation states achieved for each microstructure. The 620 °C/4 h microstructure presents a homogenous and fine precipitation state, whereas the 760 °C/2 h + 620 °C/4 h precipitates were heterogenous in size and distribution, with the presence of zones depleted in precipitation.

The TRIP effect acts during deformation by accommodating strain and besides its influence on elongation, it can be surely correlated to the energy absorption values of each microstructure under high strain rate, as reported in the literature for other steel alloys such as maraging and low carbon martensitic steels [59, 60, 61]. The 760 °C/2 h + 620 °C/4 h microstructure presented 98 % of austenite transformation and absorbed energy of 95 J evidencing the effect of this deformation mechanism on enhancing impact toughness. The results also point to the important role of this mechanism and of the austenite constituent on the strength, ductility and impact toughness synergy.

*4.3. Strength-toughness trade-off behaviour*

As evidenced by the relation between impact toughness and YS displayed in Fig. 8, both properties present the opposite tendency indicating a trade-off that usually indicates these properties are governed by different microstructural features [4]. In order to understand the conflictual nature of the relation between strength and toughness, the tensile properties presented in Fig. 12 were taken in account. Regarding the strain hardening behaviour, it can be seen from the curves exhibited in Fig. 12b the strain hardening exponent rapidly decreases for the as-built and 620 °C/4 h conditions in addition to its fluctuation and instability throughout deformation, suggesting both conditions present lower strain hardening ability than the 620 °C/8 h and 760 °C/2 h + 620 °C/4 h. The as-built microstructure would also present strain hardening at higher strains than the 620 °C/4 h microstructure. After the transition from elastic to plastic regime, both 620 °C/8 h and 760 °C/ 2h + 620 °C/4 h strain hardening exponent curves increase until a maximum value and then decrease until the necking stage. However, the 760 °C/2 h + 620 °C/4 h presents higher exponent values, indicating its higher strain hardening behaviour. Alongside its total elongation values, the 760 °C/2 h + 620 °C/4 h condition presents then higher plastic behaviour under quasi-static solicitation. The strain hardening behaviour of the PBF-LB 17–4PH alloy under quasi-static tensile test has already been clarified by other authors and associated to the strain-induced transformation of the austenite phase into martensite [13, 62, 63, 64]. The first conflictual nature of the relation between strength and toughness is indeed established by the role of the austenitic phase.

In addition to these aspects, as a precipitation-hardening alloy, the main strengthening mechanism due to the dislocation-precipitates interaction must be considered to fully understand the tensile behaviour. All the heat-treated conditions studied here present an over-aged state of precipitation for the Cu-rich precipitates, which would result in the single operation of the Orowan bypass strengthening mechanism [13, 65]. The Orowan-Ashby strengthening contribution (Δσ) can by estimated by the Equation (2) [66]:

$$\Delta\sigma = \frac{0.13Gb}{\lambda} \ln\frac{r}{b} \qquad (2)$$

In the equation, G is the shear modulus of the matrix, $\lambda$ is the interparticle spacing, $r$ is the precipitate radius and b is the Burgers vector. This would mean the strengthening magnitude would be inversely proportional to the interparticle spacing. Considering the precipitation states presented in Fig. 7, the 620 °C/4 h microstructure presents the smaller precipitate size (24 nm), as well as a fine and homogenous distribution with approximately 5.7 % of precipitate surface fraction. In contrast, the 760 °C/2 h + 620 °C/4 h condition shows a bimodal particle distribution, featuring a population of coarser particles (average size of 59 nm) and a lower precipitate surface fraction of around 3.5%, resulting in increased particle spacing. When comparing both microstructures, the 760 °C/2 h + 620 °C/4 h condition displays a larger interparticle spacing and a more heterogeneous precipitation state, with regions depleted of precipitates. These characteristics are expected to reduce the effectiveness of the Orowan bypass mechanism, thus leading to a lower yield strength (YS) in this condition. The precipitate strengthening contribution has been reported to be the most important to the material's strength regarding the other strengthening mechanisms that act in the microstructure [13].

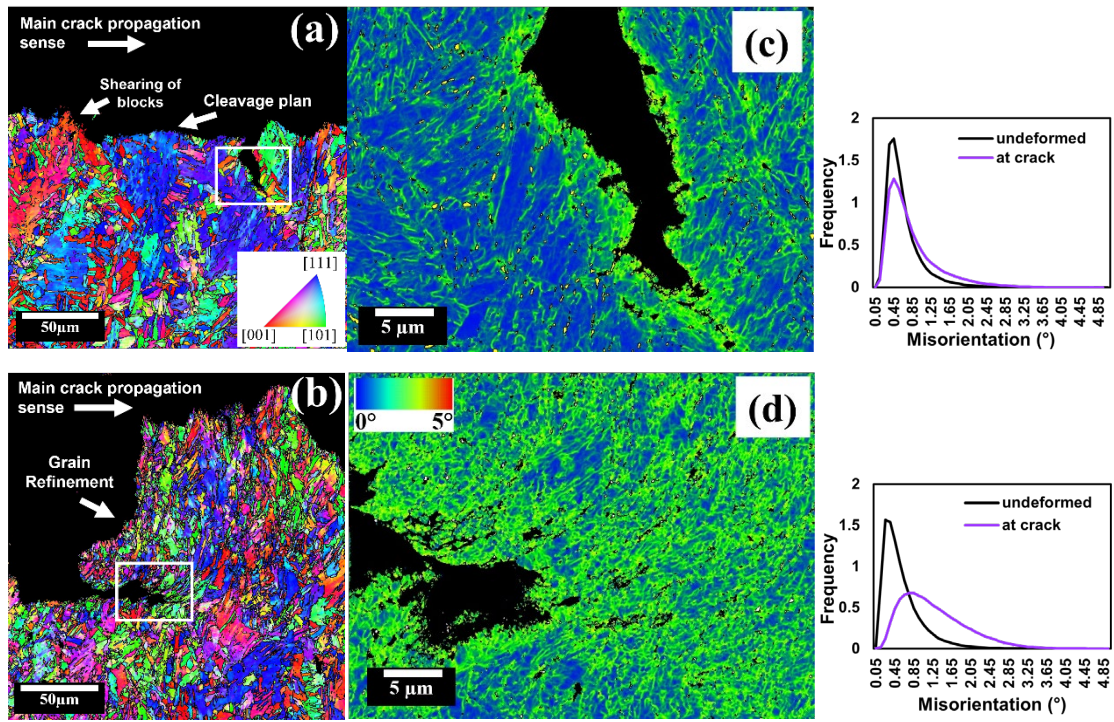

**Figure 13:** Deformed microstructure showing the main crack propagation profile and secondary cracks of the Charpy V-notch specimens. EBSD orientation maps of (a) 620 °C/4 h and (b) 760 °C/2 h + 620 °C/4 h conditions and KAM maps of (c) 620 °C/4 h and (d) 760 °C/2 h + 620 °C/4 h conditions alongside with its misorientations measures.

On the other hand, observations of the local plasticity on the crack propagation zone, as indicated by Fig. 13, displays a secondary crack in both 620 °C/4 h and 760 °C/2 h + 620 °C/4 h microstructures. It can be seen by the local misorientation around the crack and on the crack tip that the deformation of the 620 °C/4 h microstructure is localized around the crack, whereas the deformation in the 760 °C/2 h + 620 °C/4 h microstructure spreads out at further distances. This meaningful difference of the local plasticity behaviour would suggest the dislocation mobility is rather inferior in the 620 °C/4 h microstructure, where the deformation is quite restrained around the secondary crack. Even though it is difficult to assess the dislocation mobility of a specimen deformed at high strain rates, one can make assumptions regarding the main mechanism for hindering dislocation motion among the complex microstructure of the studied steel, which is the precipitation state. Taking into account the heterogenous precipitation state of the 760 °C/2 h + 620 °C/4 h microstructure, the barrier for dislocation motion would be less effective in this case, as the zones depleted in precipitates would act as free corridors enhancing dislocation mobility in this case. In opposition to this behaviour, the 620 °C/4 h and 620 °C/8 h microstructures present higher precipitate surface fraction with smaller precipitate interspace, which would be more effective barrier to dislocation motion and would result in a structure with less dislocation mobility and therefore with less local plasticity. A representation of the dislocation mobility regarding the microstructure of the 620 °C/4 h and 760 °C/2 h + 620 °C/4 h conditions is proposed in Fig. 14. The relation between precipitation state and local plasticity as a toughening aspect would therefore be contradictory to the strengthening mechanism that relies on this same feature (i.e. precipitation).

In summary, the contrasting trends in tensile strength and impact toughness discussed here relate to both the austenite volume fraction and the precipitation state of the microstructures. The results suggest that these two features govern the material's mechanical behaviour in opposing ways. A higher austenite fraction and its mechanical stability enhance impact toughness by promoting plasticity through strain-induced transformation but reduce tensile strength due to austenite's inherent softness. Conversely, a fine and homogeneous precipitation state increases tensile strength via precipitation hardening (Orowan bypass mechanism), yet restricts dislocation motion, thereby reducing local plasticity and diminishing impact toughness.

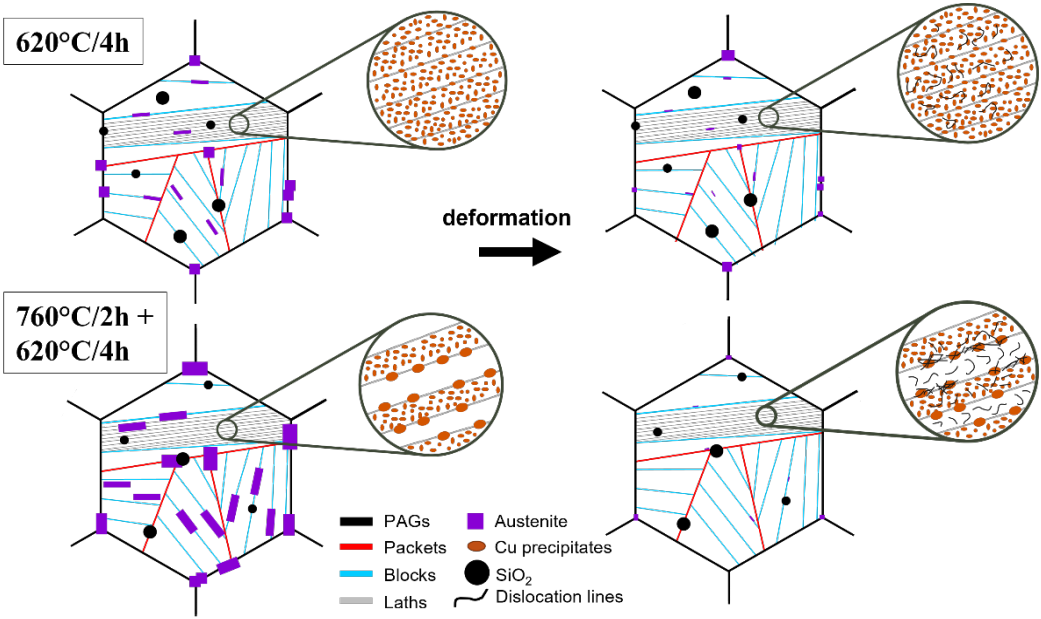

**Figure 14:** Schematic representation comparing the role of Cu-rich precipitates on the dislocation mobility in the 620 °C/4 h and 760 °C/2 h + 620 °C/4 h microstructures

## 5. Conclusions

This study correlated the complex microstructure of the PBF-LB 17-4PH, mainly hierarchic martensitic structure, soft austenitic phase, hardening nanoprecipitates and oxide inclusions, with impact toughness at the aim of validating AM for critical industrial applications. The mechanisms operating under impact loading deformation were assessed as to their influence on the impact toughness and associated microstructural feature.

1. $SiO_2$ inclusions were found to nucleate micro-cavities and micro-cracks, negatively affecting impact toughness. These particles were present at all stages of the AM process, from the initial powder to the as-built and post-heat-treated microstructures.
2. Reversed austenite contributed positively to impact toughness by promoting the TRIP effect, which enhanced local plasticity and energy absorption. Its content and stability were influenced by the heat treatments; notably, the 760 °C/2 h + 620 °C/4 h condition showed the greatest $\gamma \rightarrow \alpha'$ transformation potential. This TRIP effect also improved ductility, with the microstructure rich in austenite (760 °C/2 h + 620 °C/4 h) exhibiting superior ductility due to increased strain-assisted martensitic transformation.
3. Local plasticity was influenced by the heat treatments through their impact on dislocation mobility. The 760 °C/2 h + 620 °C/4 h condition featured heterogeneously distributed Cu-rich nanoprecipitates while retaining higher local plasticity and energy absorption capacity, resulting in improved impact toughness. However, this came at the expense of lower YS due to a reduced Orowan strengthening effect, reflecting a typical strength-toughness trade-off.

**CRediT authorship contribution statement**

**Renata de Oliveira Melo:** Writing – original draft, Conceptualization, Formal analysis, Investigation; **Jean-Bernard VOGT:** Writing – review & editing, **Eric Nivet:** Investigation; **Christophe Grosjean:** Writing – review & editing, Resources; **Eric Baustert:** Resources; **Nhu-Cuong Tran:** Resources; **Flore Villaret:** Writing – review & editing, Resources; **Jérémie Bouquerel:** Writing – review & editing, Validation, Supervision, Funding acquisition; **Gang Ji:** Writing – review & editing, Validation, Supervision, Funding acquisition

**Declaration of competing interests**

The authors declare they have no known competing financial interests or personal relationships that could have appeared to influence the work reported in this paper

**Acknowledgments**

The Chevreul Institute is thanked for its help in the development of this work through the ARCHI-CM project supported by the "Ministère de l'Enseignement Supérieur de la Recherche et de l'Innovation", the region "Hauts-de-France", the ERDF program of the European Union and the "Métropole Européenne de Lille". FIB experiments were supported by the French RENATECH network, the CPER Hauts de France project IMITECH and the Métropole Européenne de Lille.